\documentclass[aps,twocolumn,showpacs,preprintnumbers,amsmath,amssymb,showkeys]{revtex4}
\usepackage{graphicx}% Include figure files
\usepackage{dcolumn}% Align table columns on decimal point
\usepackage{bm}% bold math
\usepackage{courier}

\begin{document}

%\preprint{CRPP-PSI/2004-UT-Theor04}

\title{The self field effect on the power-law index of superconducting cables}

\author{A. Anghel}

\affiliation{Paul Scherrer Institute, CH-5232 Villigen-PSI, Switzerland}

%\altaffiliation{Also at SLS, Paul Scherrer Institute, CH-5232 Villigen PSI, Switzerland}

\date{\today}% It is always \today, today,
             %  but any date may be explicitly specified%

\begin{abstract}
It is shown that in the absence of inter-strand current redistribution the self-field effect is to always increase the power-law index of the volt-ampere characteristic and to decrease the temperature and magnetic field derivatives of the critical current line. We show that the take-off limit of a strand in a cable made of insulated strands is equal to that of a free strand due to a compensation effect between the increase of the power-law index and the decrease of the magnetic field derivative of the critical current.
\end{abstract}

%Use showkeys class option if keyword

\maketitle  %macro for all above, do not forget it!!

\section{The power-law index of a strand in a cable}
\begin{figure}[b]
	\centering
		\includegraphics[width=8cm]{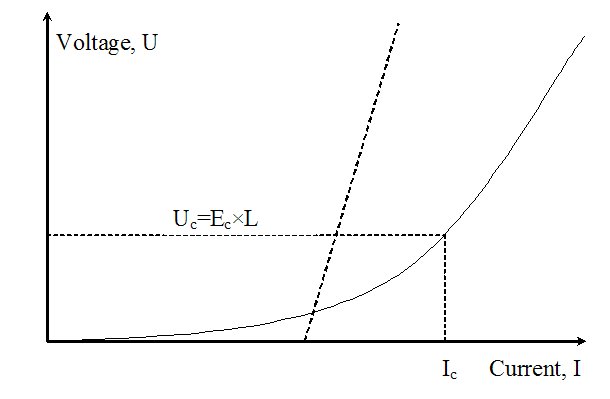}
	\caption{VAC of ideal (dashed) and real (continuous) technical superconductors. $U_c$ is a voltage criterion calculated by multiplying the electric field criterion E$_c$ with some length $L$, usually the distance between two voltage taps.}
		\label{fig1}
\end{figure}

Contrary to the statement, found usually in the text books on superconductivity, that a superconductor below its critical temperature T$_c$ conducts electricity at zero resistance, technical superconductors show a voltage drop (albeit very small) even at temperatures $T<T_c$. The volt-ampere characteristic (VAC) of this ideal superconductor (dashed-line in Fig.\ref{fig1}) would have a first range of zero resistance up to a knee point beyond which a voltage develops, linearly with the current. The knee point is identified as the critical current I$_c$. On the contrary, the VAC of a real superconductor is as shown in Fig.1 by the continuous line. The absence of the knee in the VAC in the real case request another definition of the critical current I$_c$. The most popular definition (at least in Europe and the USA) fits the smooth transition of the VAC with a power law with power exponent $n$ \cite{Walter1974}

\begin{eqnarray}
\label{eq1}
 E\left( I \right)=E_c \left( {\frac{I}{I_c }} \right)^n \nonumber \\
\nonumber \\
 I_c =I_c \left( {B,T,\varepsilon } \right)
\end{eqnarray}

with $I_c $ -the critical current as given by the scaling law (
Sommers \cite{Sommers1991} in case of Nb$_{3}$Sn, Bottura \cite{Bottura2000} for
NbTi ) \footnote{One must always distinguish carefully between the critical current as a numerical value and as a function. $I_c(B,T,\varepsilon)$ is the functional dependence while $I_c$ is a value obtained from $I_c(B,T,\varepsilon)$ for given values of $B$, $T$ and $\varepsilon$.} and $E_c $ a conveniently chosen voltage criterion, usually
0.1$\mu$~V/cm. In Eq.(\ref{eq1}) $B$ is the magnetic field, $T$
the temperature and $\varepsilon $ the strain (only for
Nb$_{3}$Sn). The power law index is $n$. 

An alternative was proposed in \cite{Dorofejev1980} and was suggested by the observed exponential increase of electrical field due to dominating thermally-activated creep over the flux flow

\begin{equation}
\label{exponential}
E=J\rho_n\exp\left(\frac{T-T_c}{T_0}+\frac{B}{B_0}+\frac{J}{J_0}\right)
\end{equation}

where: $J$ is the current density, $T_c$ the critical temperature, $\rho_n$ the normal resistivity and $T_0$, $B_0$, $J_0$ are so called \textit{grow} or \textit{increasing} fit parameters. They account for the change in VAC as a function of temperature, field and current. The relation and equivalence issues between the exponential form and the power-law VAC were investigated in \cite{Anghel2013}. Although the next calculations could be performed with the exponential form as well, we will chose to continue our analysis with the power-law functional dependence of VAC.

It is well known, although less used, that the power-law index $n$ is the
logarithmic derivative of the VAC. Indeed, if we take the logarithmic
derivative of Eq.1 with respect to the current, assuming that $I_c$ is a
constant i.e. depends only on temperature and field, we get

\begin{equation}
\label{eq2}
\frac{d\left( {\log E} \right)}{d\left( {\log I}
\right)}=\frac{I}{E}\frac{dE}{dI}=n
\end{equation}

if the VAC is described by the power-law over the whole range of
possible currents. Unfortunately, in most of the cases, the description of the
VAC by a power-law is restricted to a limited range, usually
$E\in \left[ {0.1,1} \right]\mu $V/cm and in this case we speak of
an average index in the given range. One can also define a local $n$ at
$I=I_c $. The corresponding definitions are:

\begin{equation}
\label{eq3}
\bar {n}=\left. {\frac{I_c }{E_c }\frac{dE}{dI}} \right|_{E\in \left[
{0.1,1} \right]} \mbox{ or }n_c =\left. {\frac{I_c }{E_c }\frac{dE}{dI}}
\right|_{I=I_c }
\end{equation}

Now consider the case of a cable exposed to a constant external
magnetic field $B$ in which a total current $I_t=N_s I$ flows where
$N_s $ is the number of strands in the cable and $I$ the strand
current assumed the same in all strands. The current in the cable generates a magnetic field, the self-field, which adds geometrically to the background magnetic field $B$. The resulting magnetic field in the cable cross-section becomes non-uniform. In the presence of this
self-field, the critical current of a strand in the cable $I_c $
becomes itself a function of the current in the strand. The dependence is given by

\begin{equation}
\label{eq4}
\begin{array}{l}
 I_c =I_c \left( {B\left( I \right),T} \right) \\
 B\left( {I,\varphi } \right)=\sqrt {\left( {B+B_s \left( I \right)\sin
\left( \varphi \right)} \right)^2+\left( {B_s \left( I \right)\cos \left(
\varphi \right)} \right)^2} \\
 B_s \left( I \right)=\alpha N_s I \\
 \end{array}
\end{equation}

where $\alpha $ is the geometrical self field constant depending
on the cable radius and the azimuth angle $\varphi $. For a single
round cable i.e. when the return cable is at $r=\infty $ ,
$\alpha$ is simply a geometrical constant given by

\begin{equation}
\label{eq5}
\alpha =\frac{\mu _0 }{2\pi D_c }
\end{equation}

where $D_c $ is the cable diameter.

Let us consider now for simplicity that only the maximum field
counts (peak-field hypothesis). The maximum field corresponds to
$\varphi =\pi /2 $ and from Eq. (\ref{eq4}) it is simply

\begin{equation}
\label{eq6}
B\left( I \right)=B_b +B_s(I) =B_b +\alpha N_s I
\end{equation}

a simple linear form which will be used for convenience throughout this work.

The critical current of one strand in the cable $ I_{c,cable}$ is given by the root $X$ of
the equation

\begin{equation}
\label{eq7}
X=I_{c} \left( {B_b+\alpha N_s X,T} \right)
\end{equation}

where $I_{c}(B,T)$ is the known functional dependence on field and temperature for
the superconductor (has different forms for Nb$_3$Sn and NbTi).

Now in order to calculate the index $n$ of a cable we use Eq.
(\ref{eq3}) with this new input. We have for the electric field of
a strand in the cable, which is also the electric field of the whole
cable if no current transfer takes place

\begin{equation}
\label{eq8} E_{cable} = E_{strand} = E_c \left( {\frac{I}{I_c
\left( {B_b+\alpha N_s I} \right)}} \right)^n
\end{equation}

where for convenience we do not write explicitly the
temperature dependence in $I_c $. The cable index from Eq.4 is

\begin{eqnarray}
\label{eq9} 
n_{cable} =\frac{I}{E_{cable} }\frac{dE_{cable}}{dI}=\nonumber \\
=\frac{I}{E_{cable} }nE_c \left( {\frac{I}{I_c }}\right)^{n-1}\left[ {\frac{1}{I_c
}-\frac{I}{I_c ^2}\left(\frac{dI_c }{dI}\right)}\right]=\nonumber \\
=\frac{I}{E_{cable} }nE_c \left( {\frac{I}{I_c }}
\right)^{n-1}\left[ {\frac{1}{I_c }-\frac{I}{I_c ^2}\left(\frac{dI_c
}{dB}\right)\left(\frac{dB}{dI}\right)} \right]=\nonumber \\
=n\left[ {1-\frac{I}{I_c }\left({\frac{dI_c }{dB}} \right)\alpha N_s } \right]
\end{eqnarray}

At $I=I_c$ and taking into account that $\dfrac{dI_c }{dB}<0$ we
get the simple and nice result

\begin{equation}
\label{eq10} n_{cable} =n\left[ {1+\alpha N_s \left| {\frac{dI_c
}{dB}} \right|} \right]>n
\end{equation}

i.e. in the absence of current redistribution the power-law index
in a cable is larger than the power-law index of the isolated
strands. The enhancement is proportional to the slope of the
critical current as a function of magnetic field and proportional
to the number of strands in the cable.

\section{Temperature and magnetic field slopes of critical current of the cable}
\label{subsubsec:temperature}

The critical current of a strand in a cable is different from the
critical current of a free similar strand. According to Eq.
(\ref{eq7}) the critical current of a strand in cable is
implicitly defined

\begin{equation}
\label{eq11}
I_c =\Omega \left( {B_b+\alpha N_s I_c,T } \right)
\end{equation}

where for simplicity we denote by $I_c $ the critical current of a
strand in cable $I_c \equiv I_{c,cable}$. In order to avoid
confusion we denote by $\Omega =\Omega \left( {B,T} \right)\equiv
I_{c,strand}$ the scaling relation for the free strand \footnote{One could write $I_{c,strand}=I_c(B,T)$ and $I_{c,cable}=I_c(B_b+\alpha N_s I_{c,cable},T)$ but this would be a little bit confusing.}. Although
in implicit form, using the above equation one can get a relation
between the temperature slopes of the critical current of a
strand in a cable and of a free strand. Remembering that $I_c $ is a function
of field and temperature we have from Eq. (\ref{eq11}) that

\begin{equation}
\label{eq12} \frac{\partial I_c }{\partial T}=\frac{\partial
\Omega }{\partial B}\alpha N_s \frac{\partial I_c }{\partial
T}+\frac{\partial \Omega }{\partial T}
\end{equation}

After rearranging the terms and restoring the initial notation
$\Omega \equiv I_{c,strand} $ we get finally

\begin{eqnarray}
\label{eq13} 
\left. {\frac{\partial I_c }{\partial T}}
\right|_{cable} =\frac{\left. {\dfrac{\partial I_c }{\partial T}}
\right|_{strand} }{1-\alpha N_s \left.{\dfrac{\partial I_c
}{\partial B}} \right|_{strand} }=\nonumber \\
=\frac{\left. {\dfrac{\partial I_c }{\partial T}}
\right|_{strand} }{1+\alpha N_s \left| {\dfrac{\partial I_c
}{\partial B}} \right|_{strand} }
\end{eqnarray}

We will show later by direct calculation of the derivatives that $\left. {\dfrac{\partial I_c }{\partial
B}} \right|_{strand} $ is always negative and therefore the temperature slope of the critical current in the cable is always reduced due to the self field effect.

As shown in [3] the stability limit of a conductor, expressed by
the maximum sustainable electric field before a take-off, depends
on the first derivative of the critical current with respect to
temperature. At the first sight a reduction of the temperature
slope should have consequences on the stability limit i.e. it
should increase the stability limit. We will show now that this is
not the case due to an interesting compensation effect. Indeed,
the stability limit in the simplest form i.e. for constant
power-law index is \cite{Anghel2003}

\begin{equation}
\label{eq14} E_q =\frac{hp_w }{n_{cable}}\left| {\frac{\partial
I_c }{\partial T}} \right|_{cable}^{-1}
\end{equation}

and we arrive, using Eqs. (\ref{eq13}) and (\ref{eq10}), at the
important relation

\begin{equation}
\label{eq15} E_{q,cable} =E_{q,strand}
\end{equation}

i.e. the stability limit is not affected by the self-field effect
due to the compensation effect between the increase in the
power-law index and decrease in the field slope as stated before.

The field slope is calculated similarly. Differentiating Eq.
(\ref{eq11}) with respect to $B$ we get

\begin{equation}
\label{eq16}
\frac{\partial I_c }{\partial B}=\frac{\partial \Omega }{\partial B}\left(
{1+\alpha N_s \frac{\partial I_c }{\partial B}} \right)
\end{equation}

which after some algebra manipulations becomes

\begin{equation}
\label{eq17} \left. {\frac{\partial I_c }{\partial B}}
\right|_{cable} =\frac{\left. {\dfrac{\partial I_c }{\partial B}}
\right|_{strand} }{1+\alpha N_s \left| {\dfrac{\partial I_c
}{\partial B}} \right|_{strand} }
\end{equation}

a relation similar to that of Eq. (\ref{eq13}).

\section{Comparison between $\mathrm{Nb_{3}Sn}$ and $\mathrm{NbTi}$}
\label{subsubsec:comparison}

It is instructive to compare the enhancement of the power-law
index for the two typical low temperature superconductors
Nb$_{3}$Sn and NbTi. For this purpose, the temperature and field
derivatives of the critical current for the two materials have
been calculated for a 0.8mm strand with a copper-non copper ratio
of 1.5.

For the critical line of Nb$_{3}$Sn strand we used the Sommers
\cite{Sommers1991} scaling with a typical, all-purpose set of parameters.
Two calculations were done, one at a fixed temperature of 4.5K and
a variable magnetic in the range 4-14T and one at a fixed field of
12T and variable temperature in the range 4-12K.

For the NbTi strand, the Bottura \cite{Bottura2000} scaling was used and
the parameterization was in this case a fixed temperature of 4.5K
with a magnetic field in the range 4 to 10T and a fixed field of
6T and a variable temperature in the range 4-10K.

The difference in the field ranges for the two superconductors is
understandable. One cannot compare Nb$_{3}$Sn and NbTi at the same
field because the shared part of the field ranges of the two
superconductors is at fields which are too low for Nb$_{3}$Sn and
to high for NbTi. The "standard" operating fields are: 12T for
Nb$_{3}$Sn and 6T for NbTi.

\begin{figure}[tb]
\centerline{\includegraphics[width=9cm]{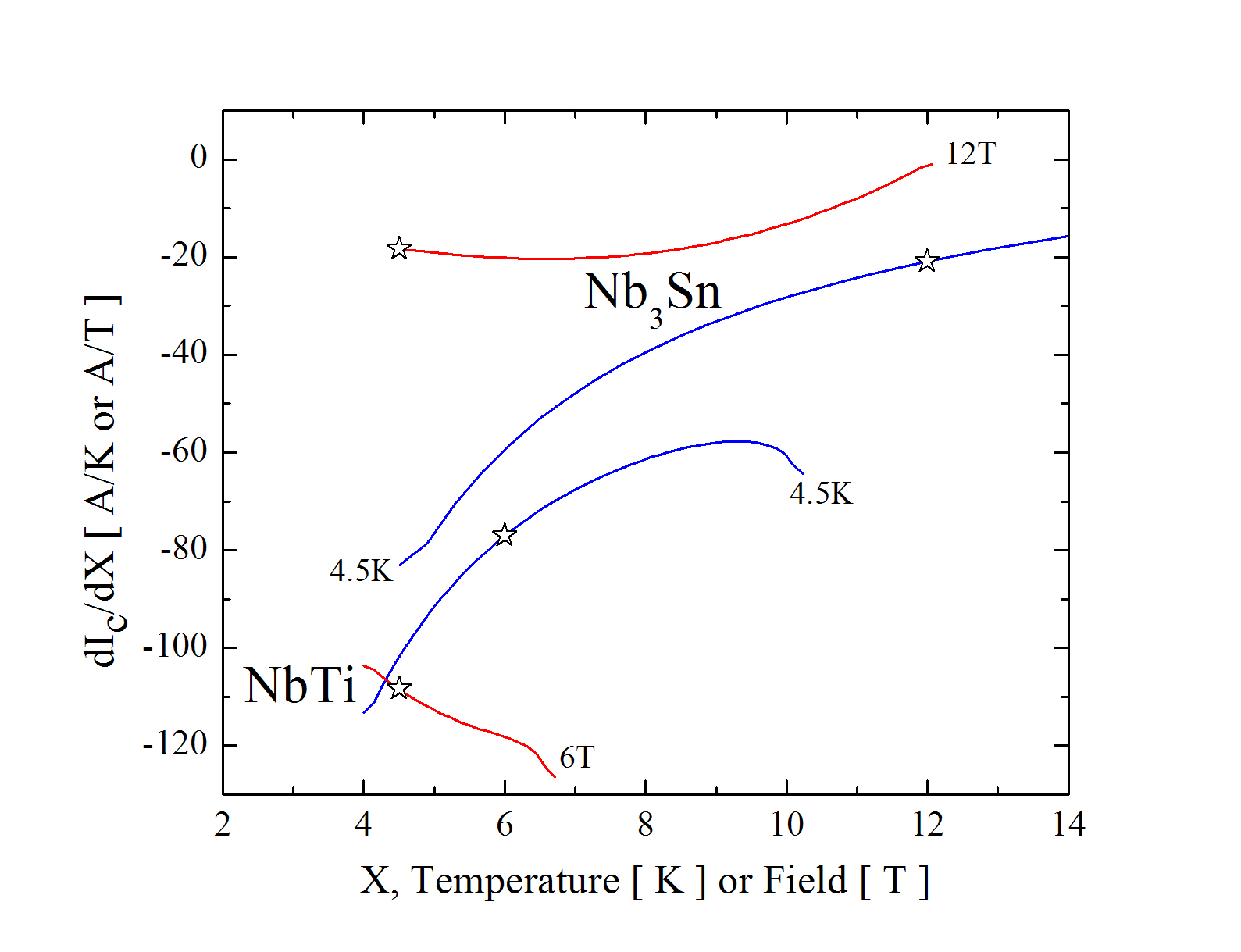}} 
\caption{(Color online) Temperature and magnetic field derivatives of critical current for Nb$_{3}$Sn and NbTi. In red, $(\partial I_c / \partial T)_B$ and in blue, $(\partial I_c / \partial B)_T $. The stars mark the reference points 4.5K and 6T for NbTi and 4.5K and 12T for Nb$_3$Sn.}
\label{fig2}
\end{figure}

The derivatives, as calculated from the corresponding scaling
relations, are shown in Fig.\ref{fig2}. Please note that in the
relation above the modulus of the derivatives appear.

It can be seen that the NbTi derivatives are larger than the
corresponding derivatives for Nb$_{3}$Sn which makes NbTi more
sensitive to the self field effect. The biggest difference is
between the temperature derivatives, around 110 A/K for NbTi and
only 20 A/K for Nb$_{3}$Sn. The difference between the field
derivatives is also large, 80A/T for NbTi and 20 A/T for
Nb$_{3}$Sn. Also noticeable is the difference between the
temperature and field derivatives of NbTi: -110 A/K and -80 A/T at
4.5K and 6T.

An important point is that the slope lines never cross the zero line i.e.
the slopes are always negative and therefore the self-field effect will
always increase the power-law index.

\section{Conclusions}

Considering a cable-in--conduit conductor as a collection of
insulated strands, relations between the power-law index and the
temperature and magnetic field slopes of the critical line of the
cable and of the free strands were deduced. The main result is
that the power-law index of a strand in the cable is moderately
increased with respect to the power-law index of a free strand.
The increase is by a factor related to the magnetic field
derivative (slope) of the strand critical current, the self field
constant and the total number of strands. The temperature and
magnetic field derivatives of the critical line in a cabled strand
are decreased by the same factor. We have shown also that due to
the compensation effect between the increase in the power-law
index and the decrease (by the same factor) of the temperature
slope of the critical current line, the stability limit of the
conductor is the same as that of a free strand.

Measurements on cable-in-conduit conductors show frequently an important
reduction of the power-law index as compared to the strand value instead of
an increase as predicted here. In view of the facts revealed here it is then
clear that the self-field cannot be the cause for this decrease. The
self-field effect on the power-law index is moderate and therefore it can be
easily removed (covered) by other mechanisms. Mechanisms that act in the
opposite direction i.e. to reduce the power-law index are: the current
redistribution (both NbTi and Nb$_{3}$Sn) and/or conductor degradation by
bending and transversal stress at the interstrand contact points (only
Nb$_{3}$Sn).

In this work we did not took into account the possible dependence of the power-law
index on the critical current. In \cite{Anghel2013} some consequences of this dependence for the stability are investigated. However, the paradigm is that this dependence is a
result of current redistribution among the strands an effect which is
beyond our goal in the present paper and is probably too complicated to be solved in the frame of
an analytical model.

\bibliography{selffield}
\bibliographystyle{unsrt}

\end{document}